\documentclass{caosp302}

\usepackage{graphicx, amsmath, amssymb, stmaryrd}

\newcommand{\zav}[1]{\left(#1\right)}

\articleNo{123}
\pubyear{2007}
\volume{38} \volnumber{3}
\firstpage{1} \received{October 1, 2007} \accepted{October 1, 2007}

\begin{document}

\hauthor{Z.\,Mikul\'a\v{s}ek {\it et al.}}

\htitle{Rotational braking of HD 37776}

\title{The record-breaking rotational braking of \linebreak
the He~strong CP star HD~37776}

\author{Z.\,Mikul\'a\v sek \inst{1,} \inst{2} \and
        J.\,Krti\v{c}ka \inst{1} \and
        J.\,Zverko \inst{3} \and
        G.\,W.\,Henry \inst{4} \and
        J.\,Jan\'\i k \inst{1} \and
        I.\,I.\,Romanyuk \inst{5} \and
        J.\,\v{Z}i\v{z}\v{n}ovsk\'y \inst{3} \and
        H.\,Bo\v zi\'c \inst{6} \and
        M.\,Zejda \inst{1} \and
        T.\,Gr\'af \inst{2} \and
        M.\,Netolick\'y \inst{1}}

\institute{ Institute of Theoretical Physics and Astrophysics,
            Masaryk University, Brno, Czech Republic, \email{mikulas@physics.muni.cz}
         \and
            J.\,Palisa Observatory and Planetarium, V\v SB TU, Ostrava, Czech Republic
         \and
            Astronomical Institute, Tatransk\'a Lomnica, Slovakia
         \and
            Tennessee State University, Nashville, Tennessee, USA
         \and
            Special Astrophysical Observatory, RAS, Nizhnyi Arkhyz, Russia
         \and
            Hvar Observatory, Zagreb University, Zagreb, Croatia
          }

\date{March 8, 2003}

\maketitle
\begin{abstract}
We study the long-term light and spectral variations in the
He-strong magnetic chemically peculiar star HD\,37776 (V901 Ori) to
search for changes of its 1.5387\,d period in 1976--2007. We analyze
all published photometric observations and spectrophotometry in the
He\,{\sc i} 4026\,\AA\--line. The data were supplemented with 506
new (\!{\it {U}})\!{\it {VB}} observations obtained during the last
2 observing seasons, 66 estimates of He\,{\sc i} equivalent widths
on 23 CFHT spectrograms and 69 on Zeeman spectrograms from the 6-m
telescope. All the 1895 particular observations have been processed
simultaneously.

We confirm the previously suspected increase of the period in
HD\,37776 which is a record-breaking among CP stars. The mean rate of
the period increase during the last 31 years is 0.541$\pm$0.020~s
per year. We interpret this ongoing period increase as the slowing
down of the star's surface rotation due to momentum loss
through events and processes in its magnetosphere.
\end{abstract}

\noindent HD\,37776 is a well-known He-strong B2\,IV star showing
rotationally modulated spectroscopic, magnetic and photometric variations with
the period 1.538675(5) days (Adelman, 1997). Mikul\'a\v{s}ek {\it et
al.}, 2007a suspected that the period is increasing.

\begin{figure}[t]
\centerline{
\includegraphics[width=0.77\textwidth,height=0.58\textwidth]{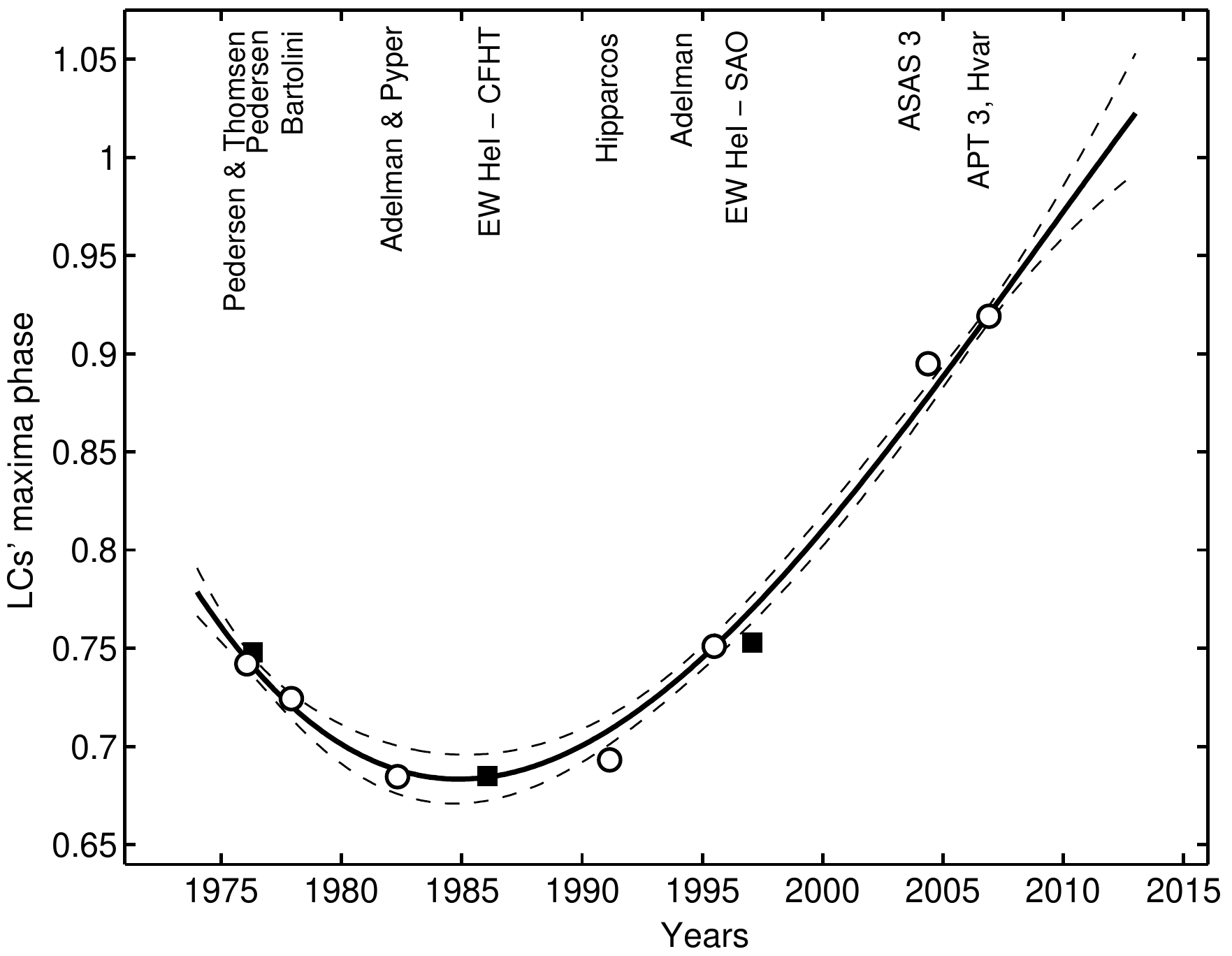}}
\caption{The O$-$C diagram of the star HD 37776, $\blacksquare$ --
spectroscopy, {\large $\circ$} -- photometry.} \label{faze}
\end{figure}

We confirm a gradual period increasing and a rate of the
increase by analyzing the extensive set of the 1895 photometric and
spectral observations obtained in ten time intervals well
distributed over the last 31 years. The times of the maxima of the
brightness, JD$_{\mathrm{max}}$, can be well approximated by a cubic formula:
\begin{equation}
\mathrm{JD_{max}}\cong
M_0+\overline{P}E+\frac{\overline{P}\overline{\dot{P}}}{2}
\zav{E^2\!-\!
\alpha_1E\!-\!\alpha_2}+\frac{\overline{P}^2\ddot{P}}{6}
\zav{E^3\!-\!\alpha_3E^2\!-\!\alpha_4E\!-\!\alpha_5}
\label{JDort}
\end{equation}
where $M_0$ is the time of the basic maximum of the light curves
near the weighted center of the measurements,
$\overline{\dot{P}}$~and $\ddot{P}$ are the mean time derivative and
the second derivative of the period, respectively, $\overline{P}$ is
the mean period. $E$ is the epoch counted from $M_0$ and
$\alpha_{1,\ldots 5}$ are constants determined by time distribution
of the data, providing the independency of individual terms in the
formula. We derived: $M_0=2\,449\,112.550(3)$,
$\overline{P}=1.5387128(8)\,\mathrm {d}$,
$\overline{\dot{P}}=1.72(6)\times10^{-8}$,
$\ddot{P}=-28(13)\times10^{-13}$\,d$^{-1}$, $\alpha_1\!=\!847.8,\
\alpha_2=\!1.029\times10^8,\ \alpha_3=-693.9,\linebreak
\alpha_4=1.297\times10^7,\ \alpha_5=1.584\times10^9 $.

There is no doubt that since 1976 the period of the star has
increased ($27\,\sigma$-certainty). Moreover, slowing down of the
rate of the period increase, (the negative~$\ddot{P})$, is not
excluded. HD\,37776 slows down its rotation or, at least, the
rotation of its surface layers. Precession and light-time effect
have been excluded. For details see Mikul\'a\v{s}ek {\it et al.},
2007b.

\acknowledgements
Grants GA~\v CR 205/06/0217, VEGA 2/6036/6 and MVTS SR~\v{C}R 01506
and 10/15 partially supported this work. {\it On-line database of mCP stars}
(Mikul\'a\v{s}ek {\it et al.} 2007c) was used. We thank D. Bohlender for
providing us with the CFHT spectra.

{}

\end{document}